\documentclass{ws-procs975x65}

\begin{document}
\begin{flushright}
Preprint INR-TH-2015-021
\end{flushright}

\title{Naturalness and Renormalization Group in the Standard Model}

\author{Grigorii B. Pivovarov}

\address{Theory Division, Institute for Nuclear Research,\\
Moscow, 117312, Russia\\
E-mail: gbpivo@ms2.inr.a.ru}



\begin{abstract}
I define a naturalness criterion formalizing the intuitive notion of naturalness discussed in the literature. After that, using $\phi^4$ as an example, I demonstrate that a theory may be natural in the $MS$-scheme and, at the same time, unnatural in the Gell-Mann--Low scheme. Finally, I discuss the prospects of using a version of the Gell-Mann--Low scheme in the Standard Model.
\end{abstract}

\keywords{Naturalness; Renormalization Group; Standard Model.}

\bodymatter

\section{Introduction}\label{intro}

By now the renormalization group functions of the Standard Model have been 
computed up to three loops \cite{Bednyakov:2014pia}. This computation is performed within 't~Hooft-Weinberg approach, which is the present-day standard. 

Below I will argue that 't~Hooft-Weinberg approach may be misleading for models with scalar fields. My argument is based on my interpretation of the naturalness criterion. 

There is some disagreement in the literature on the naturalness of the Standard Model. On the one hand, it is claimed \cite{Giudice:2013nak} that Higgs may be unnatural, and ``the multiverse offers the most plausible answer at our disposal.'' On the other hand, it seems to be claimed \cite{Boyarsky:2009ix} that the Standard Model may be self-consistent up to the Plank scale.

This disagreement may be caused by a difference in definitions of the notions used.
So, in the next section, I give my definition of the naturalness criterion.

Next, using the example of $\phi^4$ theory, I demonstrate how different renormalization schemes may lead to opposite results on the naturalness of the Standard Model.

In conclusion, I comment on the plausibility of developing an analog of the Gell-Mann--Low scheme for the Standard Model.

\section{Naturalness}

Naturalness is related to the presence of large quantum corrections to scalar masses squared \cite{Susskind:1978ms}. Formulas for these corrections may involve the ultraviolet cut-off, or some large masses of the model replacing the cut-off. In any case, the large corrections are proportional to a square of a large parameter having the dimension of mass. 

It was stressed \cite{Giudice:2013nak} that in the case when the cut-off is involved in the large corrections, it plays the role of a physical parameter. The presence of the large corrections to the scalar mass should not and does not depend on the regularization used in the computation.

The presence of the large corrections to the scalar masses is considered troublesome, because, if we want to keep the Higgs mass to the value measured by now, it implies the need for some fine-tuning between the parameters involved in the corrections.

In my view, this trouble requests an analysis of its meaning. Indeed, any computation of the scalar masses involves ultraviolet divergences. Because of this the masses can be taken as input parameters of the theory. In this case they are not computed, but just taken from an experiment. If so, how come that a computation of masses is considered troublesome? Just take the needed values from experiment, and no computation of the quantum corrections to the masses is needed regardless of their size.

Here is my answer. Indeed, it is possible to take the measurable masses as input parameters. But frequently it is computationally inconvenient. Instead of physical masses, the theoretical predictions are parametrised with  a set of input parameters related
to a particular renormalization scheme (e.g. $MS$-scheme, or momentum subtraction scheme). So, the need appears to find relationships between different sets of input parameters used in different renormalization schemes. Namely in these relationships the large quantum corrections may appear. 

There is the next question. How come that a particular change of input parameters may be considered unnatural? Mathematically, any one-to-one change of input parameters is admissible. But this is not the case in physics. A physicist implies that her/his input parameters are measurable. If a one-to-one relationship between two sets of input parameters is established within a theory, the relative accuracies with which the parameters of a set \#1 are known are translated unambiguously to relative accuracies for the parameters of the set \#2. Now imagine that we measured the parameters of a set \#1, and it appears that the parameters of a set \#2 are known because of this with relative accuracies exceeding any expectation of experimentalists. This situation seems to be unnatural. Or rather it means that we somehow overestimated the number of input parameters required in the theory, because the parameters of the set \#2 should be fine-tuned to provide the parameters of the set \#1 with the measured values. 

In high energy physics, the input parameters are determined in conjunction with a particular renormalization scheme. The latter always involves an energy scale. For example, momentum subtractions involve a momentum squared normalization point, the MS-scheme involves the dimensional regularization mass parameter, etc. The physical meaning of this parameter is that it gives the energy scale at which the input parameters are convenient to measure.

Therefore, each set of input parameters of a quantum field theory comes with the energy scale at which it should be measured. The theory predicts the dependence of the relative accuracies of the input parameters on the energy scale. This prediction is provided by the renormalization group evolution of the input parameters.

Let us consider the simplest possible model example: assume that our theory has the only input parameter---the mass squared of a scalar particle measured at the energy scale $Q^2$, $m^2(Q^2,m^2_{ph})$, where $m^2_{ph}$ is the physical mass squared. (A formal definition will be given in the next Section.) That is, assume that the exact values of all other parameters (masses and couplings) are somehow sent to us by the Gods and can be excluded from the list of the input parameters subject to measurements. Take also that  $m^2(m^2_{ph},m^2_{ph})=m^2_{ph}$, which is the definition of the physical mass squared, $m^2_{ph}$ (the mass squared measured at physical mass squared equals the physical mass squared). Under these assumptions,
\begin{equation}
\label{rescale}
\frac{\delta m^2(Q^2,m^2_{ph})}{m^2(Q^2,m^2_{ph})}=r\big(\frac{m^2_{ph}}{Q^2}\big)\frac{\delta m^2_{ph}}{m^2_{ph}},
\end{equation}
where 
\begin{equation}
\label{r-function}
r\big(\frac{m^2_{ph}}{Q^2}\big)= \frac{m^2_{ph}}{Q^2}\frac{\partial \log (m^2(Q^2,m^2_{ph})/Q^2)}{\partial (m^2_{ph}/Q^2)}.
\end{equation}

The dimensionless function $r(x)$ introduced above seems to be a useful characterization of the theory. Its value at the physical mass squared taken in the units of the energy scale $Q^2$ gives the factor transforming the relative error in the physical mass squared to the relative error in the mass squared measured at the energy scale $Q^2$.

Here is my definition of the naturalness criterion: \textit{A theory is natural if it satisfies the condition $$\lim_{x\to 0}r(x)\neq 0.$$}
Otherwise, if $$\lim_{x\to 0}r(x) = 0,$$ a theory is unnatural, because in this case one is able to improve infinitely the relative accuracy of the mass squared simply by increasing the energy scale $Q^2$ of the measurement.

It is also possible to refine this definition: I call a theory strongly unnatural if its $r(x)\sim x$ at $x\to 0$, and weakly unnatural if $r(x)\sim 1/\log (1/x^p),\, p>0$ at $x\to 0$. 

This refinement may be useful in the following way: the strongly unnatural theories become inapplicable when the reaction energies become moderately higher than the scalar masses of the theory, because the relative accuracy of the mass squared  grows in this case as $Q^2/m^2_{ph}$. This is in contrast to the weakly unnatural theories, where the accuracy grows much slower, as $\log (Q^2/m^2_{ph})^p$.

In the next Section I will try to determine the $r$-function for the $\phi^4$ theory. I will show that different answers will be obtained for different renormalization group evolution of the mass squared present in the literature. I also try to determine the source of these differences.

Concluding this section I point out that my naturalness criterion does not require the presence of a second mass in the theory whose value would be much larger when the one of the above $m^2_{ph}$. This may seem to be in contradiction with the claim \cite{Giudice:2013nak} that such a mass is needed to make the notion of naturalness meaningful. But in fact, there is no contradiction, because the second parameter of the right dimension is provided in my consideration by the energy scale of the mass measurement, $Q^2$.

\section{The $r$-function for $\phi^4$}

It follows from the definition (\ref{r-function}) that to determine the $r$-function one has to determine the running mass squared of the scalar particle, $m^2(Q^2,m_{ph})$. Remarkably, three different answers for this running are available in the literature.
Correspondingly, one has three options for the $r$-function depending on which answer for the running mass one uses.
I will first present all three options, and point out my preference after that.

But first I define the running mass squared. Its definition can be extracted from the behavior of the renormalized (UV-finite) propagator near a Euclidean momentum $Q_0$:
\begin{equation}
\label{r-mass}
D(Q^2)=\frac{Z(Q_0^2)}{Q^2+m^2(Q_0^2,m_{ph}^2)} + \mathcal{O}\big((Q^2-Q^2_0)^2\big).
\end{equation}
This defines unambiguously the field normalization constant $Z(Q_0^2)^{-\frac{1}{2}}$ and the running mass squared $m^2(Q_0^2,m_{ph}^2)$, if the propagator $D(Q^2)$ is known. The physical mass squared is defined from the equation
\begin{equation}
\label{ph-mass}
m_{ph}^2=m^2(m_{ph}^2,m_{ph}^2).
\end{equation}

So, to get the running mass squared, we need to determine the renormalized propagator. I start with the propagator renormalized within the $MS$-scheme 
\cite{Collins:1973yy,Collins:1974ce}.

The propagator renormalizaed in $MS$-scheme can be read off the above two papers:
\begin{equation}
\label{ms-prop}
D(Q^2)=\frac{1}{(Q^{2(1-\gamma_\phi)}+m_{ph}^{2(1-\gamma_\phi)})\mu^{2\gamma_\phi}}.
\end{equation}
Here $\mu$ is the mass unit of dimensional regularisation, and $\gamma_{\phi}$ is the so-called anomalous dimension of the scalar field. It was calculated in the leading order in the quartic self-coupling $g$ \cite{Collins:1974ce}:
\begin{equation}
\label{anomalous-dim}
\gamma_\Phi=\frac{g^2}{12 (16\pi^2)^2}.
\end{equation}

The definition (\ref{r-mass}) and the formula for the propagator (\ref{ms-prop}) imply that $MS$-scheme leads to the following running of the mass squared:
\begin{equation}
\label{mass-sq-ms}
m^2_{ms}(Q^2,m_{ph}^2)=Q^2\Big(\frac{m_{ph}^2}{Q^2}\Big)^{1-\gamma_\phi}.
\end{equation}
The subscript is to remind that this evolution is determined within the $MS$-scheme.

Now, using the definition (\ref{r-function}), one obtains
\begin{equation}
\label{r-ms}
r_{ms}(x)=1-\gamma_\phi,
\end{equation}
which means that the $r$-function calculated within $MS$-scheme is constant. 

I conclude that there is no naturalness problem in $\phi^4$ renormalized within the $MS$-scheme. The naturalness criterion is fulfilled.

Let us compare this with the treatment in a paper not using the $MS$-scheme, but discussing the naturalness problem \cite{Susskind:1978ms}. (This may be the first paper considering closely the naturalness problem.)

After an adaptation to the notions of the present treatment, the reasoning of this paper looks as follows. 
First, the propagator is presented in the form 
\begin{equation}
\label{s-paper}
D(Q^2)=\frac{1}{Q^2+m_{ph}^2-\Sigma(Q^2,m_{ph}^2)},
\end{equation}
where $\Sigma(Q^2,m_{ph}^2)$ is the self-energy of the scalar field. 

Next, I need some expression for the self-energy. In the original paper \cite{Susskind:1978ms} the author uses an UV cut-off and assumes that the the self-energy is proportional to this cut-off squared. Within the present reasoning, I emulate this by assuming that the leading behavior of the self-energy at large $Q^2$ is 
\begin{equation}
\label{emulation}
\Sigma(Q^2,m_{ph}^2)=-k Q^2+o(Q^2).
\end{equation}
I note here that if the propagator is renormalized with subtractions at momentum $Q$, this assumption can be justified in the leading order in the coupling, and the constant from (\ref{emulation}) turns out to be \cite{Pivovarov:2009wa}
\begin{equation}
\label{constant}
k=\gamma_\phi.
\end{equation}

Now I take (\ref{emulation}) and (\ref{constant}), neglect the term sub-leading at large 
$Q^2$, and substitute it in the definitions from the previous Section. This yields for the running mass
\begin{equation}
\label{r-mass-Susskind}
m^2_{S}(Q^2,m^2_{ph})=m^2_{ph}+\gamma_\phi Q^2,
\end{equation}
and for the $r$-function
\begin{equation}
\label{r-function-S}
r_S(x)=\frac{x}{x+\gamma_\phi}.
\end{equation}
Here the subscripts on $m^2_S$ and $r_S$ mean  Susskind's running mass squared and $r$-function.

I conclude that, within this treatment, $\phi^4$ is strongly unnatural, because $r_S(x)\sim x$.

There is a third treatment in the literature of the mass squared evolution in $\phi^4$ \cite{Pivovarov:2009wa}. It is obtained by applying a version of the Gell-Mann--Low scheme to the $\phi^4$ theory. As a result, a nonlinear equation for the renormalized propagator is obtained. It implies a running of the scalar mass squared at large $Q^2$:
\begin{equation}
\label{r-mass-GML}
m^2_{GML}(Q^2,m^2_{ph})=\frac{\gamma_\phi Q^2}{1+4\gamma_\phi\log\frac{m_{ph}^2}{Q^2}} + o\Big(\frac{Q^2}{1+4\gamma_\phi\log\frac{m_{ph}^2}{Q^2}}\Big).
\end{equation}
Here the subscript means ``Gell-Mann--Low running.''

Two remarks are in order about this result. First, it is in agreement with (\ref{r-mass-Susskind}): if the first term in its right-hand-side is expanded in powers of $\gamma_{\phi}$, the leading term will be $\gamma_{\phi}Q^2$, as in the right-hand side of (\ref{r-mass-Susskind}). Second, resummation of the leading logs of $Q^2$ has resulted in a Landau pole in the running mass squared taken in the units of $Q^2$, $m^2_{GML}/Q^2$.

Now I can determine the $r$-function corresponding to the Gell-Mann--Low running of the mass squared:
\begin{equation}
\label{r-GML}
r_{GML}(x)=\frac{-4\gamma_\phi}{1+4\gamma_\phi\log \,x}.
\end{equation}

Interpretation of this result is complicated by the presence of the Landau pole. If $Q^2$ is smaller than $Q^2_{LP}$, the momentum squared corresponding to the location of the Landau pole, the absolute value or the negative $r_{GML}$ increases with $Q^2$ to infinity. If $Q^2$ is larger than $Q^2_{LP}$, the positive $r_{GML}$ decreases to zero.

I conclude that, if I use $\phi^4$ as an effective theory at momenta below the Landau pole, there is no naturalness problem in this theory. This change with respect to the naive treatment ignoring the logarithms of $Q^2$ is due to the resummation of the leading logs of $Q^2$. 

If it is sensible to consider $\phi^4$ at momenta above the Landau pole in the mass squared, where the naturalness problem reappears? I don't know the answer. 

I take that the presence of the Landau pole in the mass squared is a 
feature of $\phi^4$, and that in a more realistic model it may be replaced with the asymptotic freedom behavior: the scalar mass in the units of the energy tends to zero as inverse logarithm of $Q^2$.

For the $r$-function that would mean a change of the sign in front of the logarithm in (\ref{r-GML}). Such  $r$-function would have the asymptotic $1/\log(1/x^p)$ at  zero $x$, where  $p$ is the coefficient by the logarithm replacing within the more realistic model the factor $-4\gamma_\phi$ of the $\phi^4$.

I conclude that resummation of the leading logarithms of $Q^2$ in the mass squared may result in a weakly unnatural behavior of mass squared. 

In further discussion I assume that the Landau pole in the mass squared is replaced with the asymptotic freedom behavior. 

The first lesson I draw from the above consideration is that the logarithms of the energy scale in the mass corrections should be taken into account. Neglecting them may result in a qualitative error in a consideration of the naturalness problem.

Next, I need to compare the Gell-Mann--Low scheme result (\ref{r-GML}) to the $MS$-scheme result (\ref{r-ms}). The results differ qualitatively. The $MS$-scheme result is that there is no naturalness problem, and Gell-Mann--Low result is that a theory with scalar fields may be weakly unnatural.

Thus, I am forced to accept that fulfillment of the naturalness criterion depends on the renormalization scheme.

Looking for a reason of this qualitative difference between the schemes I recall the claim \cite{Veltman:1980mj} that dimensionally regularized theory with quadratic divergences has a series of poles in the plane of complex dimension at $4-2/N$, where $N$ is the number of loops in a contributing Feynman diagram. This is how the quadratic divergences appear within dimensional regularization. 

The $MS$-scheme does not subtract these poles, because they are away from the physical dimension. (Notice, however, that these poles approach the physical dimension infinitely close as the number of loops grows to infinity.) 

I conclude that $MS$-scheme does not subtract the poles related to quadratic divergences, while momentum subtractions do subtract them. This is a possible reason for the qualitative difference between the $MS$-scheme and the Gell-Mann--Low scheme. 

If to chose between the two schemes, I would prefer the Gell-Mann--Low scheme, because its set of input parameters has a clear physical meaning.

The only problem with the the Gell-Mann--Low scheme is that it is not developed for non-abelian gauge theories.

So, the second lesson I draw from the above consideration is that the $MS$-scheme may be unreliable in theories with scalar particles. 

What should be done to develop an alternative to $MS$-scheme for non-sbelian gauge theories? This is discussed in the next section.

\section{Prospects of Gell-Mann--Low schemes for the Standard Model}

In view of the above, it seems to be an imperative to develop a version of the Gell-Mann--Low scheme for the Standard model.

Let me remind what such a scheme should be. The Green's functions of the Standard Model should be expressed in therms of input parameters extracted with explicitly given rules from the Green's functions themselves. The parameters will be dependent on the energy scale(s) parameterizing the way of the extraction. After that, evolution equations for the input parameters describing their dependence on the energy scale(s) should be derived from the requirement of independence of the Green's functions from the arbitrary choice of the energy scale(s).

This program has been outlined in an early paper \cite{Lee:1972fj} on renormalization of spontaneously broken gauge theories. To my knowledge it has never been realized, because the $MS$-scheme approach has always seemed after the appearance of this paper much more promising. There are papers \cite{Kraus:1997bi, Kraus:1998ud} proving the existence of momentum subtraction schemes, but no papers trying to really apply Gell-Mann--Low schemes to study the evolution of the input parameters of the Standard Model.

In my view, a Gell-Mann--Low scheme should be fully developed for the Standard Model after more than forty years of negligence. At the very least, one would obtain in this way
an alternative to the $MS$-scheme. My hope is, however, that I would prove in this way that the Standard Model is weakly unnatural, and obtain a new estimate for the scale up to which it is self-consistent. Hopefully, this new estimate will not lead to expectations of new physics at 1 TeV, as the present naive naturalness treatments do.


\end{document}